\documentclass[12pt,preprint]{aastex}

\newcommand{\HH}{H$_{160}$}
\newcommand{\JJ}{J$_{110}$}
\newcommand{\II}{I$_{775}$}
\newcommand{\et}{et al.}
\newcommand{\p}{$\pm$}

\shorttitle{Near-Infrared Properties of Faint X-rays Sources}
\shortauthors{Colbert et al.}

\begin{document}


\title{Near-Infrared Properties of Faint X-rays Sources from NICMOS Imaging in the Chandra Deep Fields}


\author{James W. Colbert, Harry I. Teplitz, and Lin Yan}
\affil{Spitzer Science Center, California Institute of Technolgy,
    Pasadena, CA 91125}


\author{Matthew A. Malkan}
\affil{University of California, Los Angeles, CA, 90095}

\and

\author{Patrick J. McCarthy}
\affil{Carnegie Observatories, Pasadena, CA 91101}



\begin{abstract}

We measure the near-infrared properties of 42 X-ray detected sources from the Chandra 
Deep Fields North and South, the majority of which lie within the NICMOS Hubble Deep Field North 
and Ultra Deep Field. We detect all 42 Chandra sources with NICMOS,
with 95\% brighter than \HH \ = 24.5. 
We find that X-ray sources are most often in the brightest and most 
massive galaxies. Neither the X-ray fluxes nor hardness ratios of the sample show 
any correlation with near-infrared flux, color or morphology. This lack of correlation indicates
there is little connection between the two emission mechanisms and is consistent with the near-infrared 
emission being dominated by starlight rather than a Seyfert non-stellar continuum.

Near-infrared X-ray sources make up roughly half of all extremely red (\JJ -\HH \ $>$ 1.4) objects 
brighter than \HH $>$ 24.5. 
These red X-ray sources have a range of hardness ratios similar to the rest of the sample, 
decreasing the likelihood of dust-obscured AGN activity as the sole explanation for their 
red color. 
Using a combination of spectroscopic and photometric redshifts, we find the 
red \JJ -\HH \ objects are at high redshifts ($z>$1.5), which we propose as the primary 
explanation for their extreme  \JJ -\HH \ color. 
Measurement of rest-wavelength absolute B magnitudes shows that 
X-ray sources are the brightest optical objects at all redshifts,
which explains their dominance of the bright end of the red  \JJ -\HH \ population. 

\end{abstract}




\section{Introduction}

The Chandra Deep Fields North and South are the deepest X-ray pointings on the sky, 
with integration times of 2 and 1 Msec, respectively. They resolve 80-90\% 
 of the hard X-ray background \citep{cam01,cow02,ros02,ale03}, 
identifying active galactic nuclei (AGN) out to redshifts greater than $z$=5 \citep{baa03}. 
Hardness ratios suggest that many faint X-ray sources are obscured AGN, 
although some of these may exist in galaxies where the bolometric 
luminosity is dominated by star formation rather than black hole accretion \citep{bau02, alb03}.  

Bright X-ray sources have well defined optical to X-ray flux ratios 
(f$_{X}$/f$_{opt}$$\sim$0.5-1), seen in both soft \citep[0.5-2 keV;][]{leh01} and hard
\citep[2-8 keV;][]{aki00} X-rays. However, fainter X-ray sources can have 
optical fluxes covering a much wider range. A large population of star-forming 
galaxies becomes detectable at X-ray depths of 1-2 Msec \citep{hor01,hor02,bab02}, 
where supernova remnants and mass transfer binaries produce the bulk of X-rays  
\citep{fab89}. These star-forming galaxies 
add low f$_{X}$/f$_{opt}$ sources to the distribution \citep{bra01,ale01,bar03}, disrupting 
the previously well-behaved optical/X-ray correlation. 
Correlations also exist between X-ray and radio \citep{bau02}, 15$\mu$m \citep{ale02, fad02}, 
and sub-mm \citep{bar01,alb03} detections. Examinations of X-ray spectral slopes and emission
lines in these sources suggest a substantial percentage of these sources are
dominated by star formation, although AGN activity can not usually be ruled out.
These X-ray sensitivities can even detect X-rays from``normal''  star-forming galaxies, 
like our Milky Way, out to $z$=0.15-0.3, which make up $\sim$12\% of the 2 Msec 
CDF-N sample \citep{hor03}.

X-ray detected sources are typically among the reddest galaxies when color is measured
from the optical to near-infrared \citep{mus00,baa01,cow01,cra01,ale01}. Roughly 10-20\% of Extremely 
Red Objects (EROs) are detected in the X-ray \citep{alb02,roc03}, only 2/3 of which are 
clearly AGN. Generally, the more optically faint an X-ray object is, the redder its optical/near-IR color 
will be \citep{ale01}, although there are some faint blue exceptions. 

Optically faint sources are blue in V-I \citep{koe02}, red in I-K \citep{ale01}, and generally 
have flat (i.e. hard) X-ray slopes. Higher redshifts may explain both the blue V-I and red I-K colors,
a result of the 4000\AA \ break and falling UV continuum shifting beyond the I-band filter at 
$z>$1 \citep{bar03}. Large X-ray luminosities are required to detect galaxies at high 
redshift, so these X-ray sources found at such distances are likely AGN, only
with optical/near-infrared fluxes that are dominated by starlight.



Studies that combine spectroscopic and photometric redshifts 
find that most faint (f$_{X}$(0.5-2 keV)$>$5$\times$10$^{-16}$ ergs cm$^{-2}$ s$^{-1}$) 
X-ray sources lie in the redshift range $z$=0.2-1.3, with a median redshift of $z\sim$1
and a long tail of sources trailing out to very high redshift ($z>$4) \citep{bar03}. 
This indicates that the bulk of optical observations to date 
have taken place in the rest-frame ultraviolet and blue. These wavelengths are notoriously susceptible to 
dust extinction and recent star formation activity, but remain fairly insensitive to the underlying mass 
distribution of the galaxy.  

Deep ground-based optical and near-infrared imaging detects roughly 85\% of these 
faint Chandra X-ray sources \citep{mus00,baa01,ros02,bar03}. Even fairly deep {\it Hubble 
Space Telescope} optical imaging can not always identify every X-ray source \citep{koe04},
indicating a need for deep, near-infrared space imaging.

We examine NICMOS imaging of 42 X-ray sources from the Chandra Deep Fields North and South. They lie mainly within the NICMOS HDF-N and and the NICMOS Ultra Deep Field (UDF), although a small subset of sources come from other publicly available NICMOS data.
We examine their colors, hardness, x-ray flux, and morphology and compare that to other NICMOS field galaxies. 
Throughout this paper we assume a $\Omega _M$=0.3,  $\Omega _{\Lambda }$=0.7, H$_o$=70 km s$^{-1}$ Mpc$^{-1}$
cosmology. All magnitudes are given in the AB magnitude system. Often, \HH , \JJ , and \II \ will be used in 
place of F160W, F110W, and F775W to simplify discussion of magnitudes, depths and colors.


\section{Observations}

We use the {\it Chandra X-ray Observatory} point source catalogs
for both the 2 Msec CDF-N \citep{ale03} and the 1 Msec CDF-S \citep{ros02}. 
We used the catalogs generated by \cite{ale03} for both fields, to ensure a consistent comparison of the two fields and to maximize the positional accuracy (median positional offset from optical detections).
The CDF-N catalog contains 503 sources over roughly 250 square arcminutes, although
the sensitivity and source counts drop away from the center. The CDF-S catalog contains 326 X-ray
sources over roughly the same area, with a similar sensitivity distribution. The CDF-N field reaches
0.5-8 keV fluxes of $\sim$10$^{-16}$ ergs s$^{-1}$ cm$^{-2}$, while the CDF-S reaches 0.5-8 keV 
fluxes $\sim$2$\times$10$^{-16}$ ergs s$^{-1}$ cm$^{-2}$. The 42 sources that lie within NICMOS 
images are listed in Table 1, labeled by the letter 'N' for north or 'S' for south, combined with 
its number from the \cite{ale03} catalogs.

The near-infrared images used for this study were obtained with NICMOS \citep{tho98} on board the {\it Hubble Space Telescope}. The data include the NICMOS HDF-N \citep{dic00}, the NICMOS UDF (HST Program 9803, PI: R. Thompson), and seven smaller fields taken in pure parallel mode\citep{mcc99} or for other unrelated science goals. 
The NICMOS HDF-N covers 7 square arcminutes of the CDF-N, reaching depths in the F160W filter of \HH =27 magnitudes ($\sim$5-$\sigma$, 0.8$\arcsec$ aperture). The NICMOS UDF also covers about 7 square arcminutes and reaches \HH $\sim$27.5. Both images were drizzled \citep{fru97} as part of their original reduction, producing images with a 
point-spread function (PSF) of FWHM $\sim$0.25 arcsec. The two fields were taken using NICMOS Camera 3 and include
 F110W (\JJ $\sim$26.5 and $\sim$27.1) images in addition to F160W. Because these fields cover only $\sim$15 square arcminutes of sky, this study could be strongly effected by cosmic variance, although we are helped by the combination of two distantly separated regions of sky.

We chose the ACS F775W filter for comparison between optical and near-infrared properties. For the NICMOS UDF, we used the F775W Ultra Deep Field image (HST Program 9978, PI: S. Beckwith), which covers the entire 
near-infrared image down to an AB magnitude of 29.5 (S/N $\sim$ 5).
For the HDF-N, we used the F775W GOODS data set \citep{gia04}, which reaches an AB magnitude of 
around 27.5. While the Hubble Deep Field F814w filter observation is deeper (F814W$\sim$28.0), it does not cover 
the entire NICMOS HDF-N and has no equivalent in the south. The majority of the X-ray detected sources 
are significantly brighter than the GOODS limit (median \II $\sim$22.2 and 95\% are brighter than \II =27.5 
limit), indicating the lesser depth is an acceptable trade-off for the filter consistency and field coverage. 
To minimize systematic errors associated with comparing galaxies at different resolutions, we convolved 
the ACS images with a Tiny Tim NICMOS PSF to match the NICMOS resolution.  

We also include NICMOS images taken in parallel to other observations \citep{mcc99} within the Chandra Deep Field 
North. This provides five additional sources from three parallel NICMOS fields, all in the CDF-N 
(1235+6208, 1235+6218, \& 1237+6206). There are presently no CDF-S NICMOS parallels that overlap 
with X-ray sources. The parallel images are not drizzled, so have slightly lower image resolution 
(PSF $\approx$ 0.3$\arcsec$), and reach depths of H$\sim$26. They lie outside the GOODS field, so we 
use the Subaru SuprimeCam Hawaii Deep Field North I-band image \citep[I$\sim$25.6 limit, 0.7$\arcsec$ seeing;][] 
{cap04} to get comparable optical data. Finally, we include two NICMOS Camera 2 images 
with one X-ray source each. Both fields were originally taken for separate science programs -- a study of 
high-z Type Ia Supernovae (HST Program 9352, PI: A. Riess) and Tully-Fisher galaxies (HST Program 7883, 
PI: N. Vogt). While shallower (\HH $\sim$25.5 and $\sim$23), they have superior resolution (PSF $\approx$ 
0.15$\arcsec$). Both lie within the GOODS fields, split between the south and north. 
None of the parallel images have corresponding F110W images. One of the Camera 2 images does have F110W data,
 although the source detected in the F110W filter is less than 5-$\sigma$ in significance.

The X-ray coordinates are generally accurate to better than 0.6$\arcsec$ as measured from 1.4 GHz radio 
(North) and R-band (South) source positions \citep{ale03}, making misidentification of the near-infrared counterpart unlikely. The median positional difference between the X-ray and NICMOS near-infrared centroids is 0.11$\arcsec$, with a standard deviation of 0.24$\arcsec$. Only one source has
an offset greater than 0.7$\arcsec$, N245 at 1.3$\arcsec$. No other near-infrared source can be 
reasonably identified with it and positional errors allow this faint X-ray source to lie in the 
outer parts (as opposed to the center) of the presently associated NICMOS galaxy.

\subsection{Magnitudes and Optical/Near-Infrared Colors}

To create a catalog of near-infrared sources, we ran SExtractor V.2.0.19 \citep{ber96}
on the F160W images, extracting objects with 9 connected pixels with signal greater than 1.5-$\sigma$
above the local background. 
For photometry, SExtractor measured a BEST magnitude, which is either a Kron elliptical aperture 
\citep{kro80} or an isophotal magnitude for blended/crowded objects. We further required 
that the flux be at least a 5-$\sigma$ detection as measured within a 0.8$\arcsec$ aperture. 
Only one of the 42 X-ray sources failed detection by this second criteria: N231. With a
\HH \  magnitude of 27.29$\pm$ 0.21, it fell just below the 5-$\sigma$ HDF-N limit. Another source, S259, was weakly detected just above the 5-$\sigma$ limit in its NICMOS Camera 2 image at H=25.34.
The other 40 sources have AB magnitudes less than \HH =24.5, much brighter than the limiting magnitudes 
for all the NICMOS fields examined ($\sim$26-27.5 magnitudes).  
  
To measure colors, we ran SExtractor on the F110W and convolved F775W images using the source
positions from the F160W images and measured the flux within a 0.8$\arcsec$ aperture. 
The only exception to this method is for the ground-based 
Hawaii Deep Field North data, which has a much lower spatial resolution. Rather than blur the NICMOS
data by a factor of 3-4, we compared the SExtractor BEST magnitudes. This approach should be considered 
less accurate, but we see no systematic difference in I-\HH \ color between these parallel objects and 
the rest of the NICMOS sample.

\section{Asymmetry and Concentration}

To study the morphologies of the NICMOS sample, we use the CAS system 
\citep{coa03, cob03, con00}, which provides indices for concentration,
asymmetry, and clumpiness. 
We measured these indices using CAS software written for IRAF\footnotemark (C. Conselice, 
private communication). The resolution of NICMOS Camera 3 images is poorly suited
for measuring clumpiness at high redshifts, as the clumpiness index is most vulnerable to 
decreasing resolution, so we will exclude it from further discussion. 
While the asymmetry index is also sensitive to resolution, the effect is smaller and more predictable 
\citep{cob03}. More importantly, resolution effects always produce smaller asymmetries, so large values 
of the index remain a reliable indicator of strong asymmetry. 
A full decomposition of the light profile was rejected as a morphology measurement 
because it is too dependent on a priori assumed morphologies, which are known to become more irregular at 
high redshift \citep{abr96,con04} and because the NICMOS Camera 3 
resolution is too low to allow a robust decomposition of a large percentage
of the sources (median NICMOS object FWHM $\sim$ 1.5$\times$ PSF). 

\footnotetext{IRAF is distributed by NOAO, which is operated by AURA, Inc., under contract to the NSF.}

Simulations by 
\cite{con00} demonstrate that one can obtain reliable asymmetry values with a S/N=50, although
S/N=100 is recommended for best results. Seven of the 42 X-ray sources
do not meet the minimum (S/N=50) threshold and are excluded from morphological analysis. Of 
the rest, only two fall between S/N=100 and 50. Visual inspection of source morphologies is
consistent with their measured concentration and asymmerty parameters.  

We measure the concentration and asymmetry indices using five blank sky regions and average the results. This method follows \cite{coa03}, who find that asymmetry can vary depending on the area selected for blank sky. 

The X-ray sources have mostly symmetric NICMOS morphologies, with a distribution similar to those of 
non-X-ray emitting NICMOS sources (see Figure 1). Two objects stand out as strikingly asymmetric 
(N272 \& N287) and therefore possible merging systems. 
This result is in agreement with
the optical result of \cite{gro03} that the distribution of asymmetry indices for X-ray sources is indistinguishable from that for field galaxies. 
Nucleus-dominated sources are very symmetric, so a bright nuclear source could hide a more disturbed 
asymmetric morphology. Unfortunately, most spiral and elliptical galaxies are also highly symmetric, 
making asymmetry a poor discriminant for nuclear bright AGN.

\cite{gro03} also find significantly higher concentrations among X-ray sources. This differs from \cite{hor03}, 
who found no significant difference in concentrations for X-ray sources, although they only examined 
the optically brightest sources (R $<$ 22 compared to I $<$ 23.5). The near-infrared sources appear 
to be slightly more concentrated compared to the main sample (median C$_{Field}$=2.7 versus C$_{X-ray}$=2.9), 
but not as strongly as seen in the \cite{gro03} optical sample. Sources with bright nuclei should
generally lead to higher concentrations, although \cite{gro03} cautions that may not always be the 
case.

To investigate these concentration differences further, we measured 
the morphologies of both the 
X-ray and field sources in the original, non-convolved \II \ images. 
We find that the distribution of field \II \ concentrations are 
essentially identical to that seen in \HH \ images. The agreement between the 
\II \ and \HH \ field concentrations indicates
that image resolution is unlikely the cause for the difference between 
\II \ and \HH \ X-ray source concentrations. This is further supported
by the simulations of \cite{cob03}, which demonstrate that reliable 
concentration indices can be measured in NICMOS images out to $z$=3.

The \II \ concentrations for the X-ray detected
sample show a 0.2-0.3 increase in typical concentration compared to
\HH . The difference between the X-ray and field sample concentrations
measured in \II \ is therefore the same as that seen by \cite{gro03}.   
We conclude that X-ray sources are in fact slightly less
concentrated in near-infrared light, possibly an indication of increased
contribution by more distributed stellar light.

\section{Correlation With X-ray Properties}

The X-ray full-band flux (0.5-8 keV) shows no correlation with either the \HH \ magnitude (see Figure 2a), 
the \JJ -\HH \ color (see Figure 2b) or the morphology of the NICMOS image, all producing Pearson correlation 
coefficients (PCC) less than 0.12. 
Visually, Figure 2a appears to suggest an anti-correlation between \HH \ and X-ray 
flux (PCC=-0.11), but this is caused by the lack of near-infrared bright galaxies with large X-ray fluxes
within the relatively small area covered.
These near-infrared bright, X-ray weak objects are all at low redshifts, where the X-rays appear to come from 
star formation. 
If the X-ray full-band flux shows any correlation it is with \II - \HH \ color (PCC=0.17; see Figure 2c),
where larger X-ray fluxes appear to produce redder colors, although with a large scatter. 
If \HH \ is uncorrelated with X-ray flux, then an \II - \HH \ correlation would imply that \II \ grows 
fainter with X-ray flux, which is opposite to what is generally 
observed \citep{leh01}. However, most studies have been done at much brighter X-ray fluxes than those reached
in the Chandra Deep Fields. Examining the CDF-N, \cite{bar03} shows that the R magnitude to X-ray flux 
(both 0.5-2 keV and 2-8 keV) correlation breaks down at these faint X-ray flux levels, 
as nearby, quiescent star-forming galaxies make up a larger percentage of X-ray sources. 

Using the soft-band (0.5-2 keV) X-ray detections has little effect on these results, with the possible 
effect of marginally strengthening the weak X-ray flux and \II - \HH \ color correlation (PCC=0.24). 
The hard-band (0.5-2 keV) X-ray detections show weaker correlations than either the full-band or soft-band 
fluxes, although the number of hard-band X-ray detections is smaller (22), making robust correlation
measurements more difficult. 
None of the weak correlations discussed at any X-ray bandwith could be considered 
even moderately convincing and all measurements are consistent with no correlations whatsoever.

This lack of correlation indicates the source of the X-rays does not make much contribution to the near-IR continuum in most of these objects. Studies of bright, nearby AGNs show a correlation between X-ray and near-infrared flux \citep{ede86,kot92}, although the measurements concentrate on
the central regions. Most of the NICMOS X-ray sources lie at high redshifts, so a better rest-wavelength 
comparison would be the optical/X-ray correlation seen in bright X-ray sources \citep{aki00,leh01}, where the optical fluxes encompass the whole galaxy. The faint X-ray sample could be different, either because the rest-frame optical emission from the active central Seyfert nuclei is heavily obscured by dust, or perhaps because most of the X-ray emission is coming from extended starbursts rather than an active nucleus. 

All the reddest objects (\JJ -\HH \ and  \II - \HH ) have been detected in hard-band, while there is a blue 
population of faint X-ray faint objects detected only in soft-band. This does not, however, indicate a correlation
between color and X-ray hardness. The hardness ratio (HR), defined as (H-S)/(H+S) where H is hard-band (2-8 keV) 
counts and S is soft-band (0.5-2 keV) counts \citep{ros02}, shows no correlation with near-infrared flux or 
color (see Figure 3), although the smaller number of hard-band detections makes these correlation measurements 
less significant. It appears that the fainter near-infrared fluxes and redder near-infrared colors can not be 
explained entirely by obscured active galactic nuclei, where a dusty torus or other obscuration is absorbing both 
optical/near-infrared and X-ray light. Most likely X-ray sources at these faint flux levels are a heterogeneous 
group, composed of both obscured and non-obscured AGN, and star-forming galaxies, ranging from starburst to 
``normal'' Milky Way levels of star formation.

\section{Near-Infrared Faint Objects}

We detect all 42 X-ray sources in the NICMOS images and only one of these is detected 
with less than a 5-$\sigma$ significance. Previous attempts at imaging optically faint 
X-ray sources using ultra-deep VLT/ISAAC imaging (K$\sim$24) \citep{koe04,yan03}, still left a 
handful of sources either unidentified or only marginally detected. 
Two of the faint NICMOS detections from this work would have been undetectable in such ground-based studies, 
although neither is as optically faint as the undetected sample from \cite{koe04}. It should be kept 
in mind, however, that this study covers only two small regions of sky and therefore cosmic variance 
will have a large effect on all the results, particularly those concerning extremely red objects, 
which have been measured to vary a great deal (factors of 3-7) from field to field \citep{sar01,lab03}. 

The faintest near-infrared source, N231, has relatively blue colors: \JJ -\HH = -0.11$\pm$0.4 and 
I$_{814}$-\HH = 0.29$\pm$0.3. Its X-ray flux is weak, but is detected at roughly twice
the faint detection limit of the CDF-N sample. It also does not have a hard X-ray 
index (HR $<$ -0.2). It is much too faint to produce any reliable morphology indices, but visual 
inspection of the image suggests a diffuse, non-compact shape. Template fitting suggests
a redshift around $z$=1, but such fits should be treated cautiously considering 
the large (10-40\%) photometric errors involved. 

The other faint source, S259, has colors neither particularly red nor blue (\JJ -\HH \  $>$ 0.45; 
\II -\HH =1.65$\pm$0.12) and a simlar hardness ratio (-0.2). Its shape appears compact, but the
signal-to-noise is too low for quantitative morphological measurement. 

There are nearly three magnitudes between the near-infrared flux of N231 and the majority of the X-ray sources. The next faintest object, S259, is still two magnitudes brighter than N231. This suggests that N231, and perhaps S259 as well, are different types of object or in a different phase than the rest of the X-ray sample.   

Regardless of the origin of these two objects, the percentage of the total 
X-ray population that are near-infrared weak is not large, representing only 5\% of the
sources. The rest of the sample (H$<$24.5) can be detected using the deepest ground-based near-infrared 
photometry \citep[ex. the VLT FIRES project achieved a 5-$\sigma$ limit of H$\approx$25.5,][]{lab03}. 

\section{Extremely Red NICMOS Sources}

Several studies have shown that faint X-ray sources are, on average, redder in optical/near-infrared color 
as near-infrared flux decreases \citep{ale01,bar03}. 
We see this relationship continue at fainter near-infrared magnitudes, down to \HH $\approx$ 24.5,
below which we do not detect X-ray sources (with two exceptions as noted above). 
Figure 4 shows \II -\HH \ color versus \HH \ magnitude. 

There are 14 Extremely Red Objects (EROs, defined as \II -\HH $>$3.1; equivalent to \II -\HH $>$4 in Vega magnitudes) down to \HH =24.5 in the NICMOS HDF-N and UDF, of which 3 (21\%) are detected X-ray sources (1 in HDF-N, 2 in UDF). Considering the small numbers involved, this is consistent with both the 8-11\% found by \cite{roc03} and \cite{alb02} at comparable depths (K$<$24), and the 21\% found by \cite{alb02} in a shallower ERO survey (K$<$22). We note that the X-ray flux of the HDF-N ERO is one of the faintest in our sample and would therefore not have been detected in the present CDF-S or the earlier 1 Msec CDF-N surveys. The two UDF EROs, S233 and S246, were previously noted in the \cite{roc03} sample.

The brighter near-infrared X-ray sources have \JJ -\HH \ colors 
typical of the faint galaxies generally detected in NICMOS images (see Figure 5). 
However, a large fraction of the faint near-infrared X-ray sources (\HH $>$21.5) are significantly 
redder than a typical field galaxy. If we define extremely red \JJ -\HH \  objects as any source with  
\JJ -\HH \ $>$ 1.4, the predicted color of an elliptical galaxy at a redshift of $z$=2, 5 of 14 (36\%) 
faint near-infrared X-ray sources are extremely red. While a small number of sources, 
the large gap in \JJ -\HH \ color may indicate that there are two subpopulations of faint, X-ray sources. 
There is a similar trend for near-infrared objects seen in \cite{cow01}, where all three lensed 
X-ray sources observed had extremely red colors consistent with evolved galaxies at redshifts $z>$1.4,
and in \cite{yan03}, which found extremely red near-infrared colors for 4 of its 6 optically unidentified 
X-ray sources.

Not only are many faint near-infrared X-ray sources extremely red in \JJ - \HH , but a high percentage 
of all red \JJ - \HH \ sources are detected in X-rays.
Red \JJ -\HH \ colors appear to select X-ray sources at a higher rate than optical/near-infrared 
colors (i.e. EROs). 
If we select all extremely red  \JJ -\HH \ objects in the HDF-N and UDF down to 
our \HH =24.5 limit, 5 out of 10 are X-ray sources. Less strict \JJ -\HH \ color cut-offs produce slightly
lower percentages, but do not change the general conclusion that a large percentage of 
red \JJ -\HH \ objects are detected in X-rays.  
Down to the faintest near-infrared sources (\HH $\sim$24.5), the reddest object at almost every 
magnitude interval is an X-ray source. 


\subsection{Interpretation of the Reddest Sources}


Red \JJ -\HH \ sources may be heavily obscured AGN. 
Studies of low redshift, red J-K sources from 2MASS, for instance, produce a high
percentage of dust-extincted AGN \citep[$\sim$80\% ][]{mar03}.  
However, the red NICMOS X-ray sources do not, in general, have the 
hard X-ray slope that one might expect from a heavily extincted AGN (see Figure 3). 

A soft X-ray slope does not completely rule out strong nuclear 
dust obscuration. \cite{wil02}, for instance, find that dusty, red J-K AGNs from 2MASS 
do not have hard X-ray ratios, which they attribute to a decoupling of near-infrared reddening and
X-ray absorption, possibly from large dust grains or different dust-coverage over the optical/IR and 
X-ray-emitting regions. On the other hand, they find that the 
reddest objects were also those faintest in X-rays, while four of the reddest NICMOS sources are 
among the X-ray brightest in the sample, both in flux and absolute luminosity. Also, the 
majority of known red J-K AGN lie at much lower redshifts than the NICMOS faint X-ray sources,
so that the measured near-infrared fluxes are sampling different rest wavelength regimes.


Another possibility is that the red \JJ -\HH \  sources are at redshifts slightly above 1.5, 
where the UV slope has moved into the F110W bandpass.  \cite{cow01}, for instance, suggested 
that many optically faint Chandra sources seemed to lie in high redshift evolved galaxies, producing
extreme near-infrared colors.
We assembled redshifts for the X-ray 
sources in the HDF-N and the UDF, using published spectroscopic redshifts where available 
\citep{low97, coh99, bar03, szo04}. For the remaining seven sources (3 in HDF-N, 
4 in UDF) we used published photometric redshifts \citep{fer99,wol04}, supplemented
by redshifts produced using a Bayesian photometric redshift code \citep{ben00}. 
We plot the resulting redshifts versus \JJ -\HH \ color in Figure 6a.   

Of the six red \JJ -\HH \ objects, all are at redshifts greater than 1.5. Two have spectroscopic
redshifts: S245, a class 2 QSO at $z$=3.064 \citep{szo04} and N220, a radio galaxy identified at a 
redshift of $z$=4.424\footnotemark \citep{bra01,wad99}.  
\footnotetext{There is reason to doubt the N220 redshift, as it based
on a Ly$\alpha$ detection an arcsecond from the radio source and its optical counterpart \citep{bar00}.
Its slightly blue B-V colors also suggest that it is at a $z<3.5$ \citep{cri04}. However, the
lack of any detection of emission lines suggest the source is at $z>$1.5 \citep{bar00}, i.e.
at a high redshift. Therefore we will continue to use the $z$=4.424 redshift for this paper, as  
it makes little qualitative difference to our results exactly how high the redshift is.}
Probably the most well studied of these red objects is N267, with a 
HDF-N photometric redshift of 2.75 \citep{fer99}. This object (also known as NICMOS J123651.74+621221.4)
is the second reddest source in the NICMOS HDF-N
and has been previously noted for its color, x-ray, radio and mid-infrared properties 
\citep{dic00,hor00,aus99,ric98}. 
The three remaining sources, all in the UDF, have photometric
redshifts of 1.61, 1.74, \& 3.02. The red X-ray sources are small and compact in shape, 
except for S246\footnotemark .

\footnotetext{S246 is larger and appears to be a disk galaxy with a bright, red core. 
This extended image may indicate that the $z$=1.61 photometric redshift is incorrect. COMBO-17 gives a photometric redshift of $z$=0.27 \citep{wol04}. However, optical spectroscopy of this object and the compact $z$=1.74 S233 source produced no redshift \citep{szo04}, leading us to believe the higher redshifts are likely correct.} 

The reddest sources have 0.5-2 keV luminosities greater than 10$^{42}$
ergs s$^{-1}$ (Figure 6b), except for S243, which is only detected in hard X-rays. 
For comparison, the most X-ray luminous starburst known, NGC 3256 \citep{mor99}, 
has a 0.5-2 keV luminosity of $\sim$3$\times$10$^{41}$ ergs s$^{-1}$. 
More typical starbursts, like M82, 
generally produce $\sim$10$^{40}$ ergs s$^{-1}$ \citep{gri00}. Even accounting for an 
increase of X-ray luminous starbursts with redshifts, it is apparent the majority of the red 
sources must be mainly powered by AGN activity. 

The extreme red color of these high redshift X-ray sources 
results from steeply falling ultraviolet flux shortward of the 4000 
\AA \ break moving into the NICMOS filters. 
An unobscured AGN spectral energy distribution should not drop that rapidly in the ultraviolet, 
indicating that the rest-frame optical/ultraviolet 
light could be dominated by starlight.
This possibility is supported by the lack of correlation between near-infrared flux and 
X-ray flux, suggesting that they come from different sources. There is also 
no trend of concentration with either near-infrared or X-ray luminosity, 
although the X-ray sources are marginally more 
centrally concentrated than field galaxies. 
However, simulations adding point-source flux to HST galaxy profiles indicate point sources 
have only a small effect on concentration ($\sim$+0.1) and in some cases could actually cause 
a decrease in concentration \citep{gro03}.

If we accept that the explanation for the red X-ray sources is their high redshift,
and not their AGN or dust properties, then we must explain why they make up such a large percentage
of the faint, red objects. 
Any moderately red high-redshift galaxy would also
produce these near-infrared colors. One answer is that X-ray sources are 
the brightest galaxies at high redshifts. Non-X-ray emitting galaxies would be expected to make 
up a larger percentage of red sources at fainter magnitudes, possibly explaining the red sources
below \HH =24.5 in Figure 4.  
 
\subsection{The Brightest Galaxies}

To examine the possibility that high redshift X-ray sources reside in bright galaxies 
further, we plot absolute B magnitude (0.45 $\mu$m) versus redshift \citep{fer99} in Figure 7. 
The absolute magnitudes were calculated by
interpolating between flux densities measured in the filters before 
applying both a (1+$z$) K-correction and the distance modulus. 
For redshifts placing the rest wavelength for B 
beyond the measured filters, we extrapolated from the last two filters. The possible 
$z$=4.424 object (N220) was excluded altogether.

Figure 7 shows that the X-ray sources are not only the brightest objects at 
high redshift, but 
tend to be the brightest sources at {\it all} redshifts. This  even though many
 of the sources have near-infrared flux dominated by starlight, not black hole
accretion. At low redshift and low X-ray luminosity, where star formation is likely the dominant
source of X-rays, bright near-infrared galaxies would be expected. But we also find X-ray sources in the brightest galaxies 
at moderate to high redshifts, where their large X-ray luminosities make them almost certainly 
AGN. A similar examination of absolute R-band (0.7 $\mu$m) covers a lower redshift range ($z$=0-2), but also
identifies the X-ray sources as the brightest and, considering the longer wavelengths being
observed, most massive galaxies. This is in agreement with \citep{bar03}, who found large numbers of
very luminous ($>$M$^{\star }$) galaxies out to high redshifts among their X-ray detections, 
covering a much larger area than our sample, but to shallower near-infrared depths. 

We should be careful applying our general conclusion that AGNs are found in massive galaxies to
all individual X-ray sources. These faint X-ray sources are likely drawn from
three separate groups. There are those powered mainly by star formation, only visible at
low redshifts, with different ratios of X-ray to optical flux depending on strength of starburst
and quantity of older stars. Then there are the unobscured, Type I AGN, which are bright at all 
wavelengths from X-ray through infrared, with the energy produced from black hole accretion generally 
dominating over star light, except, perhaps, in the near-infrared. Only three sources are
confirmed broad-line AGN\footnotemark , although we should keep in mind the  small total numbers and
likely large cosmic variance.
The third likely group of X-ray sources are the Type II AGNs. 
The dust obscuring the central engines of 
these AGN can reprocess the bulk of optical radiation produced by the AGN into the infrared, 
possibly leaving only the optical light generated in stars. The lack of correlation between X-ray flux 
and near-infrared flux or concentration is consistent with obscured AGN making up a substantial fraction 
of X-ray sources.

\footnotetext{N59,S242,S254}

We should also keep in mind the two faint NICMOS detections, which are almost certainly low 
stellar mass galaxies. Their moderate to blue colors indicate that they are not heavily 
dust-obscured or at extremely high ($z>$6) redshifts. Assuming their X-rays come from AGN, 
black hole accretion would be the source of the near-infrared light as well, leaving very 
little power for any star light. The overall spectral energy distribution of these two sources
are consistent with that of a Seyfert nucleus \citep{ede86}.
However, these low near-infrared flux galaxies are clearly the exception, rather than the rule.  

\section{Summary}

We examine the near-infrared properties of 42 Chandra and NICMOS detected sources from the Hubble 
Deep Field North and the Ultra Deep Field. All of the Chandra sources were detected with NICMOS,
with only two sources fainter than \HH \ = 24.5.

We confirm in the near-infrared the result of \cite{gro03}, 
that there is no significant difference between the asymmetries of X-ray sources and typical 
field galaxies. X-ray emitting galaxies have marginally greater near-infrared concentrations when compared
to non-X-ray field galaxies, although the difference is smaller than that found by 
\cite{gro03}. 

Roughly 40\% of the faintest near-infrared Chandra sources (\HH $>$ 21.5) are extremely red, 
with \JJ -\HH \ $>$ 1.4. These red \JJ -\HH \ X-ray 
sources make up half of red \JJ -\HH \ objects down to \HH \ $<$ 24.5, more than twice the 
rate found for optical/near-infrared EROs. 
Using available spectroscopic and photometric redshifts, we find all the red \JJ -\HH \ sources
are at high redshift ($z$=1.61-4.424), with X-ray luminosities indicative of AGN. 
Their red colors come from falling rest-frame ultraviolet flux shortward of the 4000\AA \ break 
moving into the NICMOS filters at redshifts of $z>$1.5. 
This explanation for their red color implies that the optical/near-infrared
continuum is dominated by star light. Supporting evidence for this theory is that we find no correlation 
between near-infrared and X-ray flux or between X-ray hardness ratio and near-infrared color.

The X-ray sources in our study were found to be the optically brightest objects at all redshifts, 
indicating a strong preference of AGN for the optically most luminous galaxies. This trend holds at 
all optical magnitudes, including the longer rest-wavelengths, indicating that these X-ray detected
AGN also reside in the most massive galaxies.

This work demonstrates that NICMOS can be a powerful tool for identifying high redshift AGN. 
The current conclusions depend on only six red \JJ -\HH \ X-ray detections, 
four of which come from the UDF. Future studies covering much larger areas are required to test and 
confirm these results.
Because most X-ray sources are reasonably bright in the near-infrared, they should
be observable in relatively shallow NICMOS surveys. The NICMOS Parallels Survey \citep{mcc99}, 
for instance, covers hundreds of square arcminutes to the necessary depths.
The deepest ground-based surveys (ex. VLT/ISAAC GOODS South, P.I. C. Cesarsky),
should also be able to detect the majority of these red X-ray sources. 
Increasing the sample of red sources in well studied fields,
such as the Chandra Deep Fields, will allow us to utilize the
multi-observatory data now becoming available.  For example, 
the ultradeep Spitzer Space Telescope mid-infrared imaging \citep{dic04}
will further constrain the contribution of dust obscuration
and AGN emission. Such understanding will be crucial for interpretion and
analysis of the large number of red galaxies being found in deep 
infrared surveys.


\acknowledgments

We would like to acknowledge Chris Conselice for his input and providing us with his 
CAS code. We are also grateful to the anonymous referee for a thoughtful report. 
We wish to thank Mark Dickinson for his help with the NICMOS HDF-N.
We thank the staff of STScI, and in particular Galina Soutchkova,
for all their work and assistance. This research was supported by STScI grant GO-9865.

\clearpage


\begin{figure}
\epsscale{0.6}
\plotone{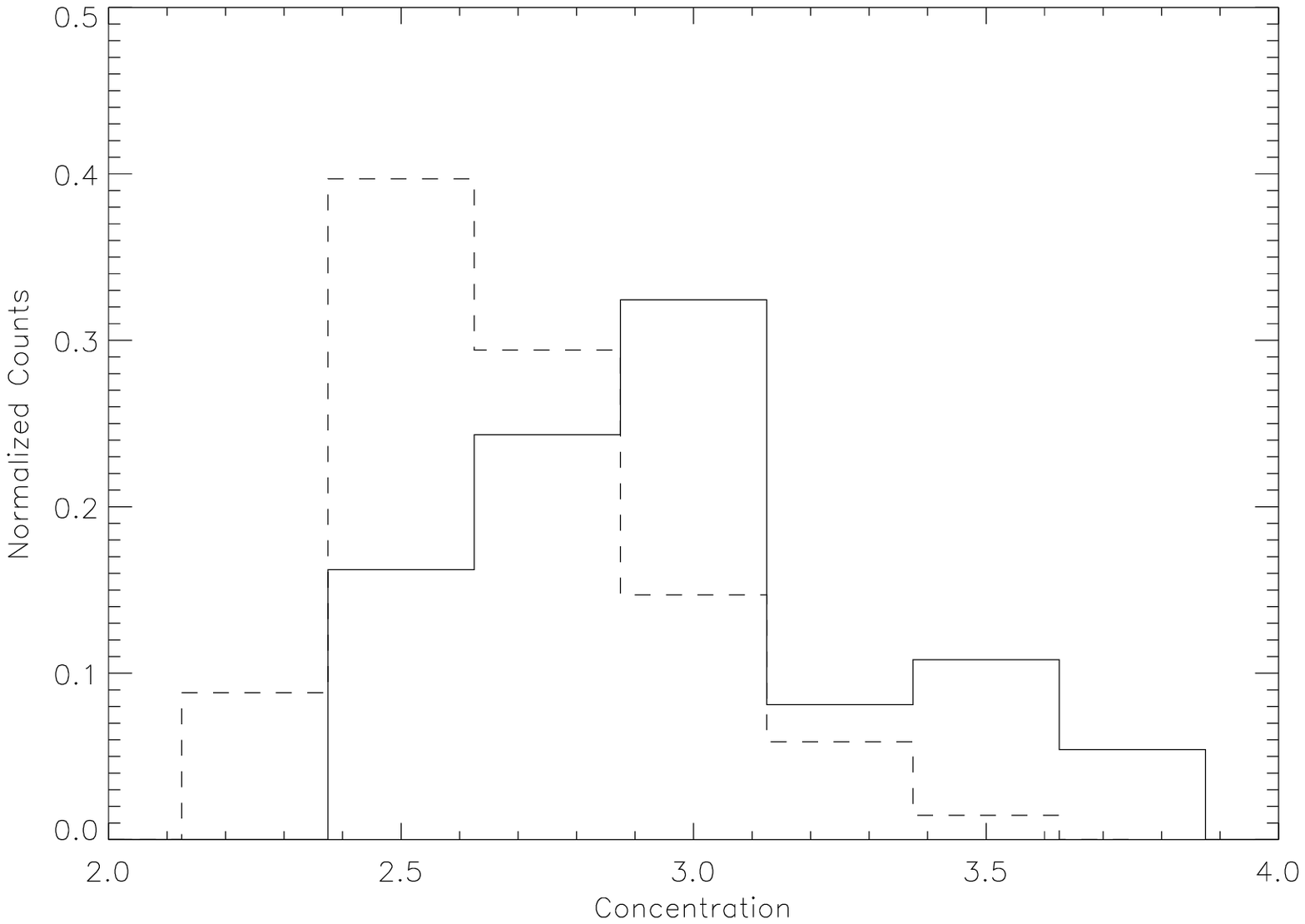}
\plotone{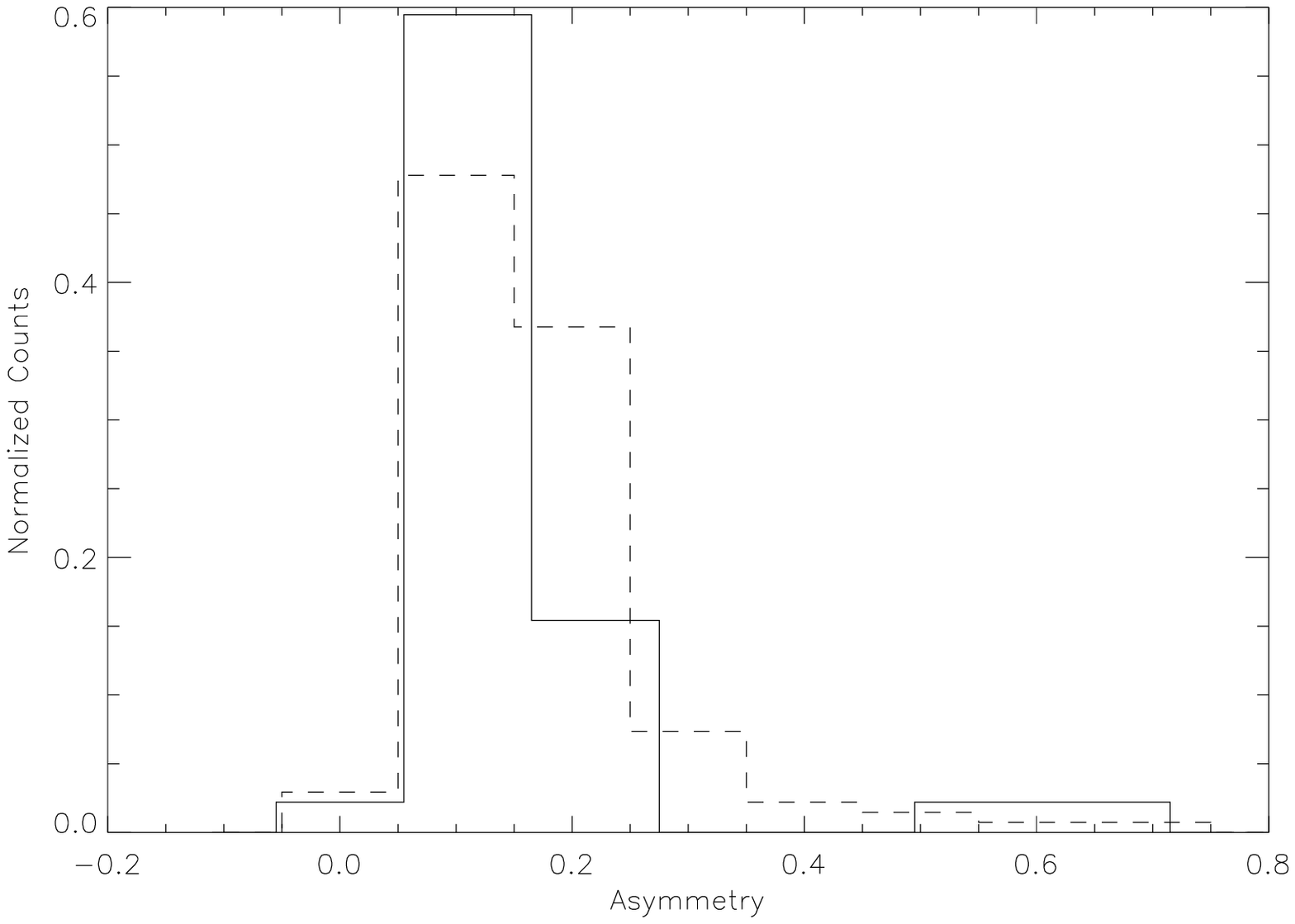}
\caption{Histograms of concentration (top) and asymmetry (bottom) for the NICMOS detected X-ray
sources (solid line) and for the NICMOS UDF and NICMOS HDF-N field galaxies (dashed line). 
Both distributions have been normalized to unity.} 
\end{figure}

\begin{figure}
\epsscale{0.5}
\plotone{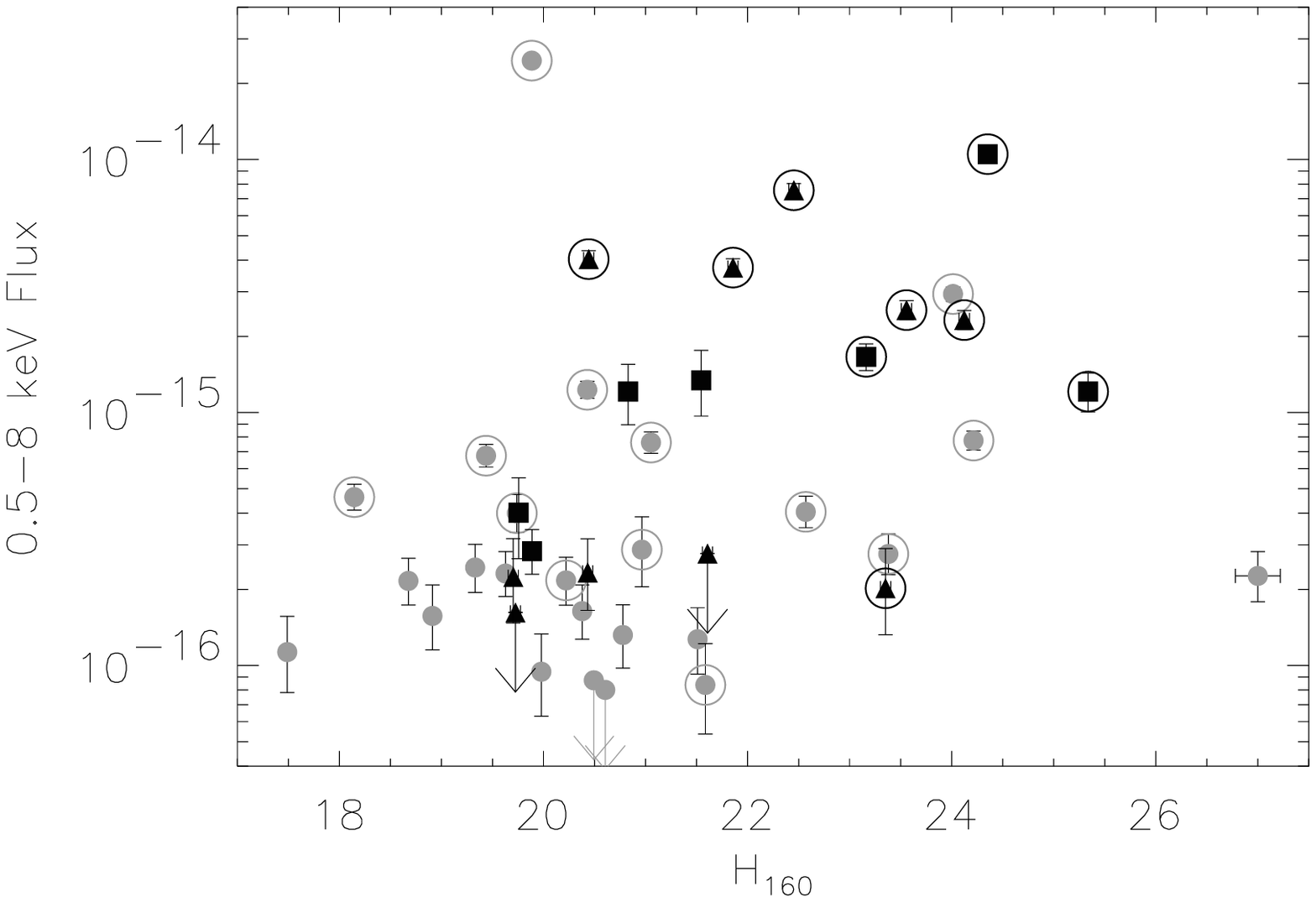}
\plotone{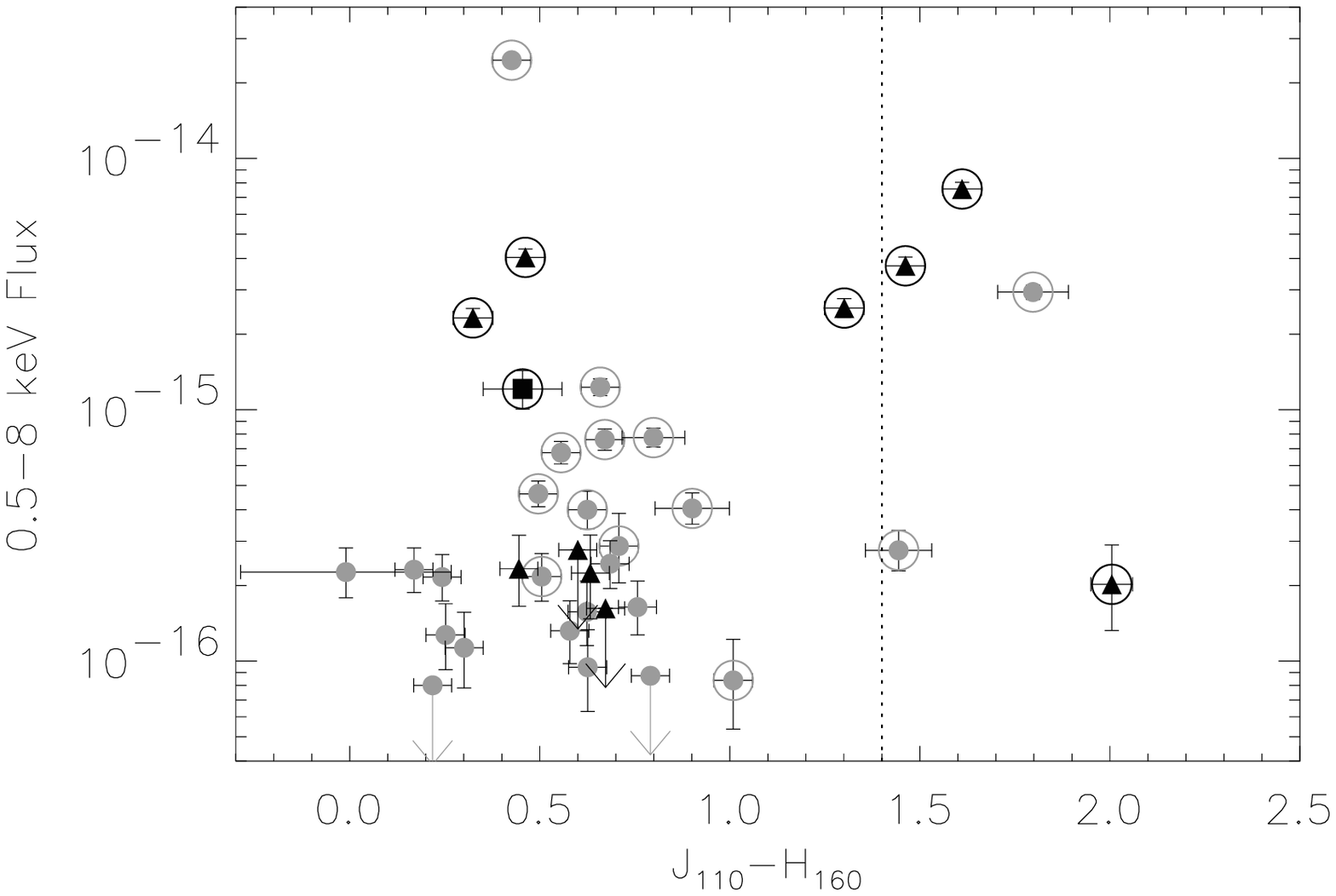}
\plotone{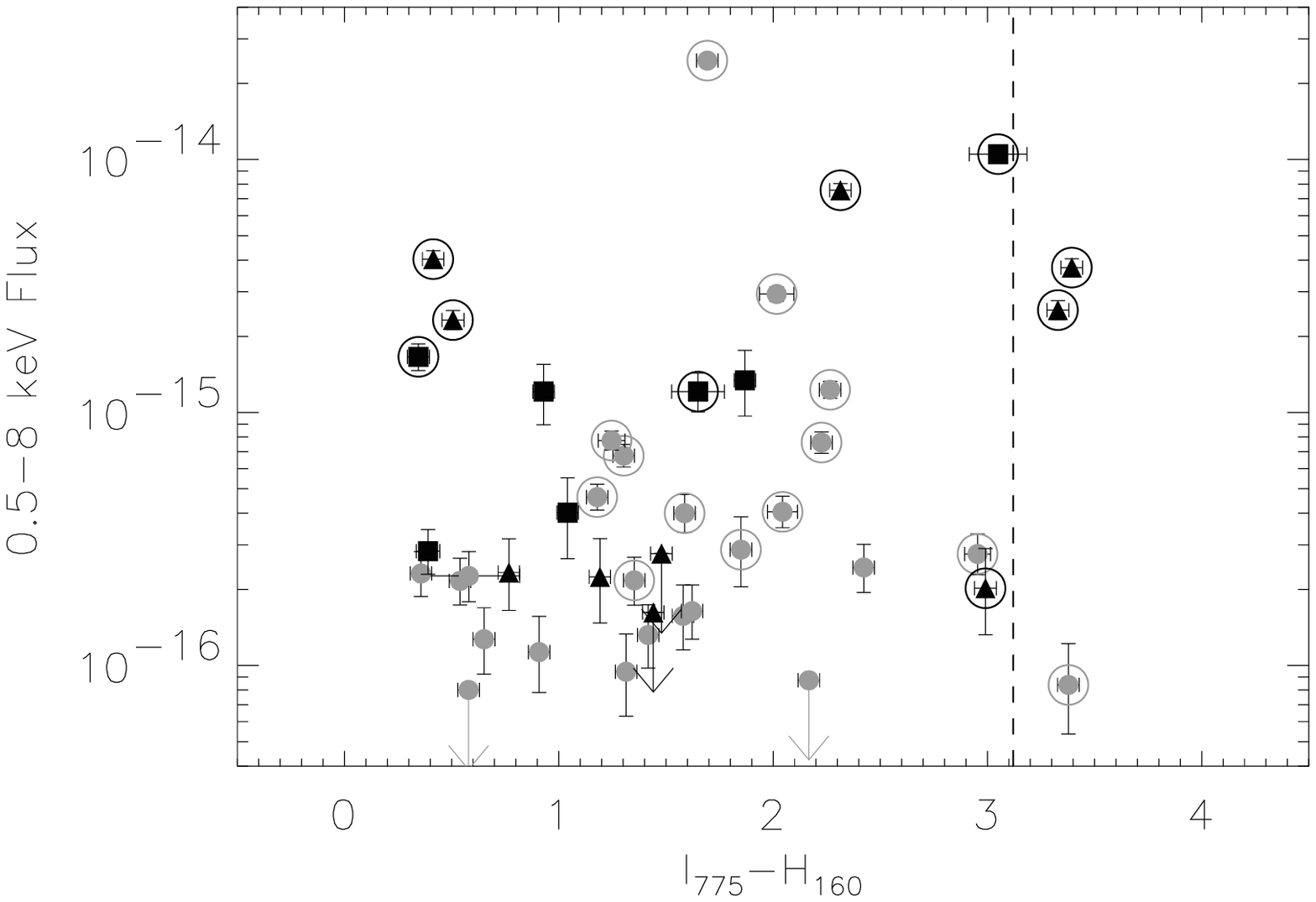}
\caption{a) 0.5-8 keV Full-band fluxes versus \HH \ magnitudes for the NICMOS sample. Solid circles
are from the NICMOS HDF-N, solid squares from the NICMOS parallels, and solid triangles from
the NICMOS UDF. Downward pointing arrows are upper limits (Faint sources can be detected in
soft-band but not in the full-band). All hard-band sources have been circled. b) 0.5-8 keV Full-band 
fluxes versus \JJ - \HH . c) 0.5-8 keV Full-band fluxes versus \II - \HH .}
\end{figure}

\begin{figure}
\epsscale{1.0}
\plotone{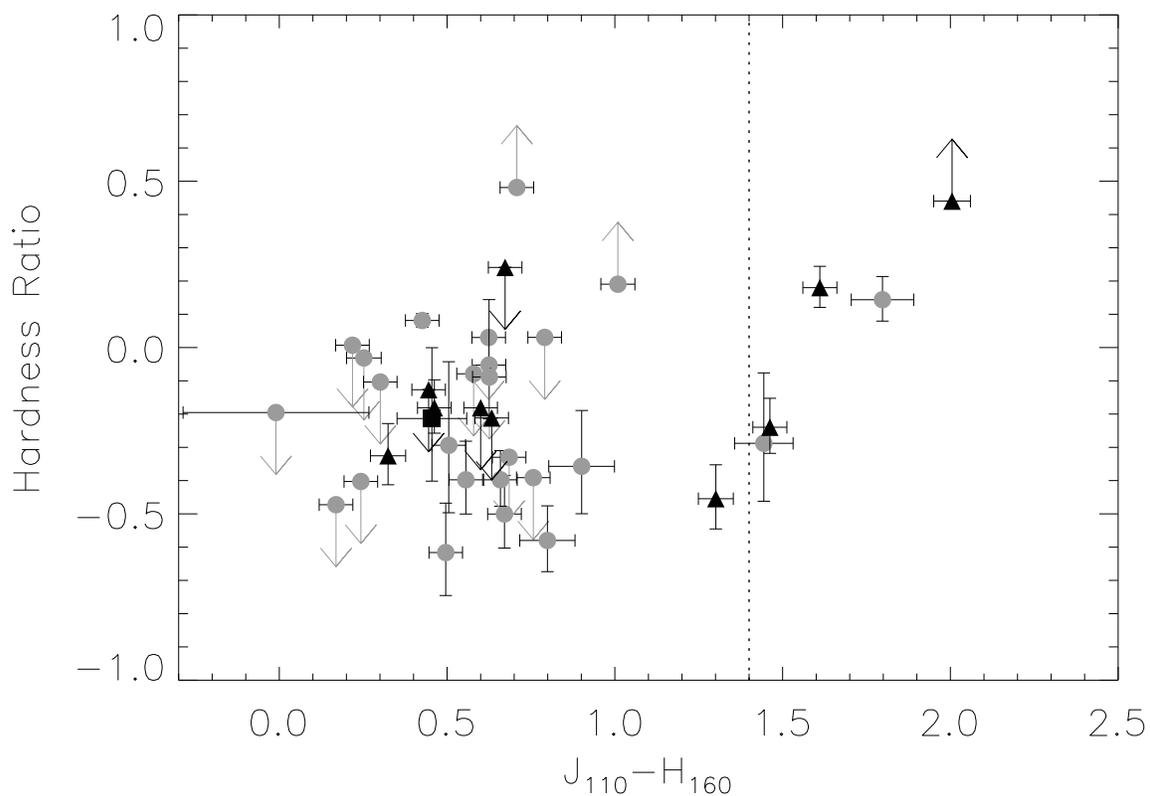}
\caption{Plot of X-ray hardness ratio (HR) versus \JJ -\HH \ color. Objects from the HDF-N are plotted with
solid circles, while those from the UDF are plotted with solid triangles. Soft-band only detections are plotted as upper limits with downward pointing arrows and hard-band only detections are plotted as lower limits with upward pointing arrows. The dividing line for extremely red \JJ - \HH \ objects is drawn as a dotted line.}
\end{figure}
 
\begin{figure}
\plotone{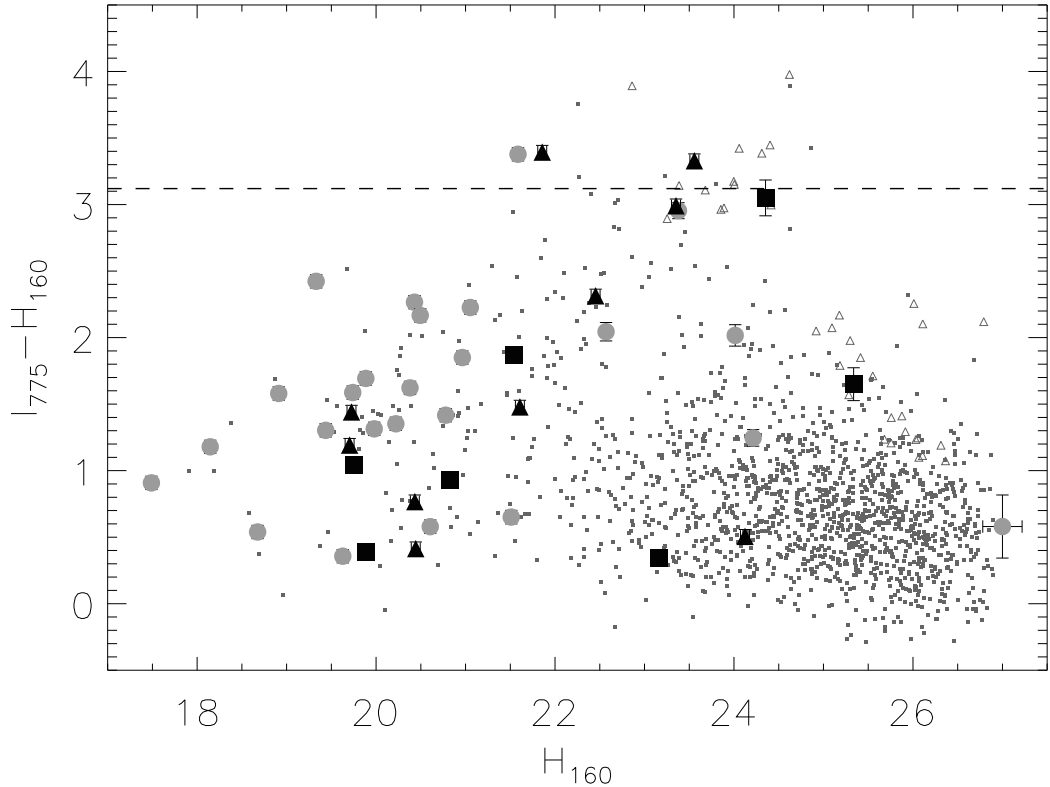}
\caption{\II-\HH \ versus \HH \ color magnitude diagram for the Chandra sources.  Solid circles
are from the NICMOS HDF-N, solid squares from the NICMOS parallels, and solid triangles from
the NICMOS UDF. For the CDF-N NICMOS parallels, I-band is from the Subaru Hawaii Deep Field North 
instead of \II . Over-plotted as single points are all
sources from the NICMOS UDF and NICMOS HDF-N. Upper limits are plotted as open, upward facing triangles.
For most sources, the error bars are smaller than the symbol size.
The dividing line for EROs (\II -\HH $>$3.1) is drawn as a dashed line.}
\end{figure}

\begin{figure}
\plotone{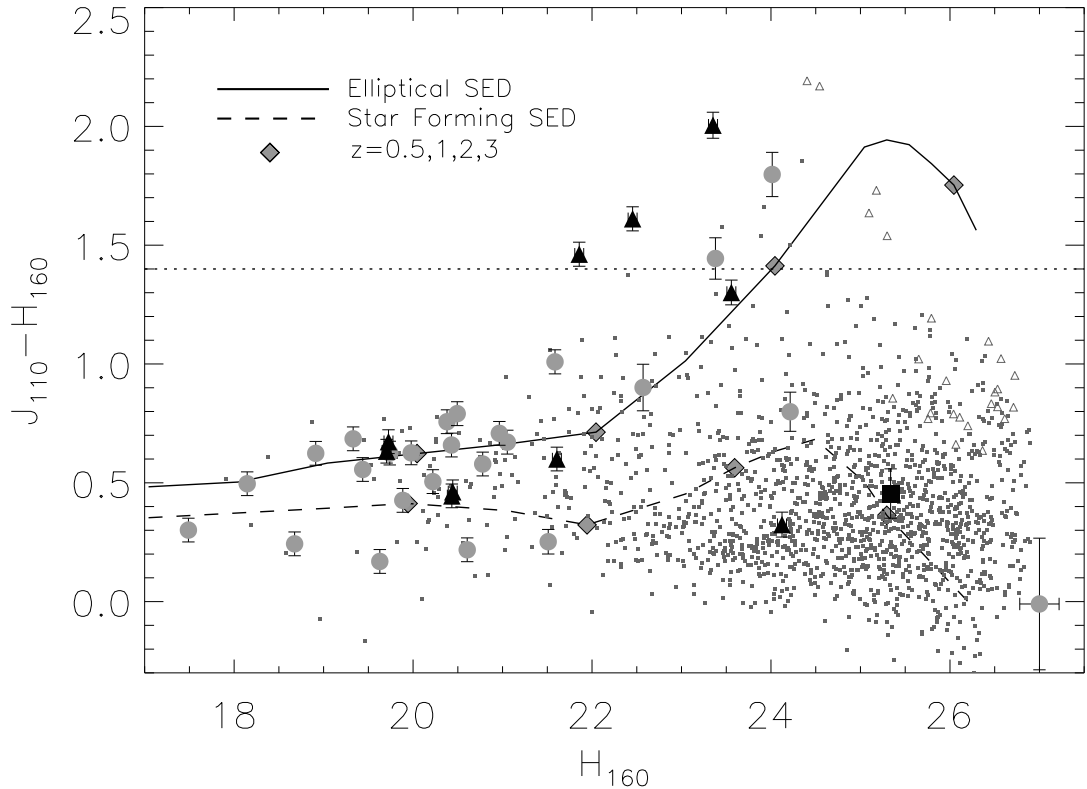}
\caption{\JJ -\HH \ versus \HH \ color magnitude diagram for the Chandra sources. 
Objects from the HDF-N are plotted with solid circles, while those from the UDF are plotted with solid triangles. 
Over-plotted as single points are
sources from the NICMOS UDF and NICMOS HDF-N down to a magnitude of F160W=26.5 ($\sim$10-$\sigma$
for the HDF-N). 
Also plotted are two model L$^*$ galaxies, where the dashed line is a spiral, while the
solid line is for an elliptical galaxy. The large black diamonds represent $z$=0.5, 1, 2, and 3 
along each model line.
The dividing line for extremely red \JJ - \HH \ objects is drawn as a dotted line.}
\end{figure}

\begin{figure}
\epsscale{0.6}
\plotone{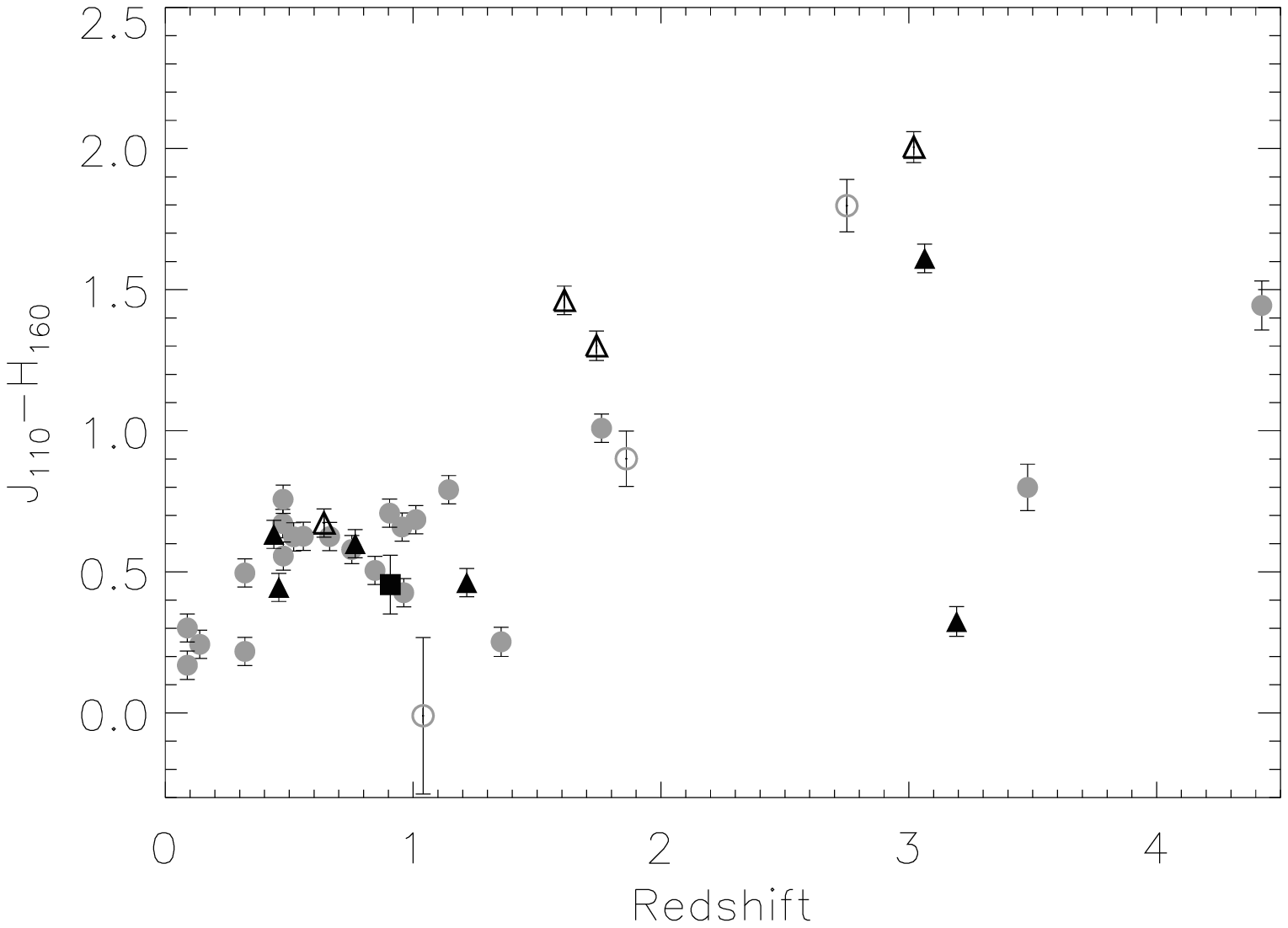}
\plotone{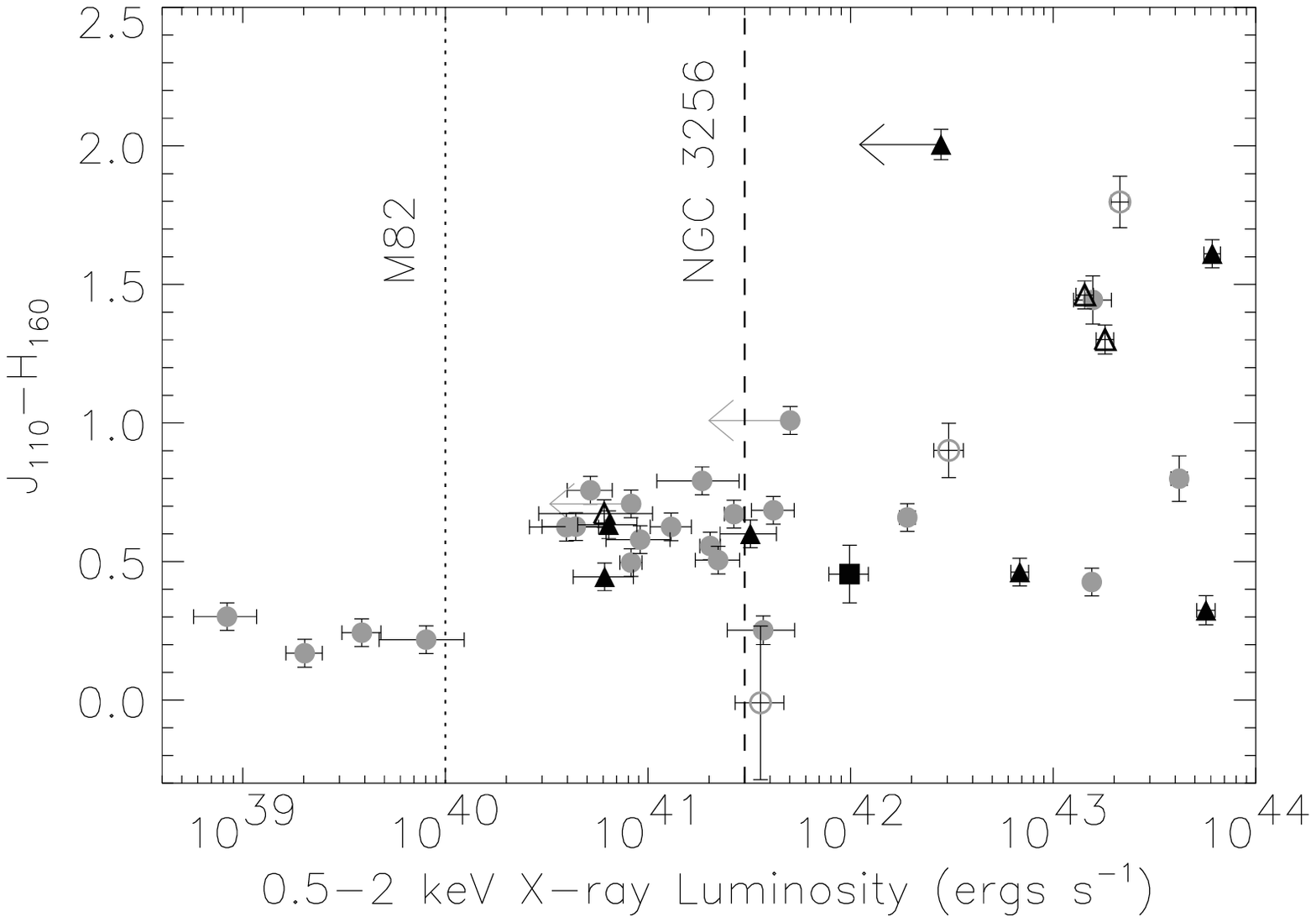}
\caption{Top: Plot of redshift versus \JJ -\HH \ color for our sample. Objects from the HDF-N 
are plotted with circles, while those from the UDF are triangles.   Sources with photometric redshifts
are plotted with open symbols, while those with spectroscopic redshifts use solid symbols.
Bottom: Plot of X-ray luminosity versus \JJ -\HH \ color for our sample. Symbols
are same as previous plot.} 
\end{figure}   

\begin{figure}
\epsscale{1.0}
\plotone{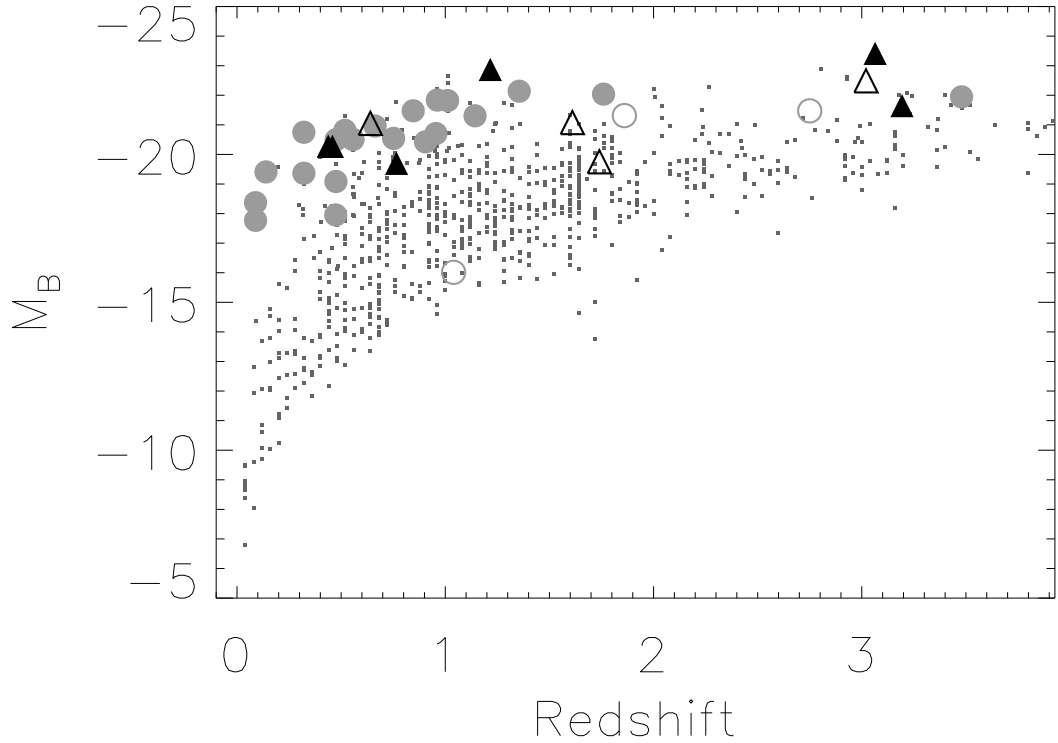}
\caption{Absolute B magnitude (AB) versus redshift. Objects from the HDF-N 
are plotted with circles, while those from the UDF are plotted with triangles. Sources with photometric redshifts
are plotted with open symbols, while those with spectroscopic redshifts use solid symbols. 
Over-plotted are all sources
in the HDF-N as points, using photometric redshifts taken from \cite{fer99}.} 
\end{figure}


\begin{deluxetable}{lllcccccccc}
\rotate
\tabletypesize{\scriptsize}
\tablecaption{NICMOS Detected Chandra Deep Field Sources} 
\tablehead{
\colhead{Obj.} & 
\colhead{RA} & 
\colhead{Dec} & 
\colhead{\HH } & 
\colhead{\JJ } & 
\colhead{\II }  & 
\colhead{$z$} & 
\colhead{C}  & 
\colhead{A} & 
\colhead{F$_{X}$\tablenotemark{a}} & 
\colhead{HR} \\
\colhead{ID\#} &
\colhead{(J2000)} & 
\colhead{(J2000)} &
\colhead{Mag} &
\colhead{Mag} &
\colhead{Mag} & & & &
\colhead{0.5-8 keV} &
}
\startdata
 N50 & 12:35:56.32 & 62:08:03.2   & 21.55 & n/a & 23.41 & 1.15 & 2.84\p 0.19 & 0.153\p 0.002 & 3.83 & $<$ -0.07 \\
 N59 & 12:36:0.41 & 62:19:25.2  & 23.16 & n/a & 23.51 & 3.19 & 2.57\p 0.25 & 0.083\p 0.007 & 4.47 & -0.32 \\
 N69 & 12:36:5.35 & 62:19:32.8  & 19.76 & n/a & 20.80 & 0.520 & 2.69\p 0.16 & 0.095\p 0.001 & 1.21 & $<$-0.01 \\
 N203 & 12:36:38.93 & 62:12:57.3 & 20.49 & 21.28 & 22.66 & 1.143 & 2.56\p 0.10 & 0.148\p 0.007 & 0.254 & $<$0.03\\
 N207 & 12:36:39.57 & 62:12:30.4 & 24.21 & 25.0 & 25.46 & 3.479 & n/a & n/a & 3.79 & -0.58 \\
 N211 & 12:36:39.92 & 62:12:49.9 & 20.22 & 20.73 & 21.57 & 0.846 & 2.94\p 0.15 & 0.096\p 0.016 & 0.641 & -0.29\\
 N218 & 12:36:41.81 & 62:11:32.1 & 19.63 & 19.80 & 19.99 & 0.089 & 2.91\p 0.10 & 0.199\p 0.005 & 1.02 & $<$-0.47 \\
 N220 & 12:36:42.11 & 62:13:31.6 & 23.38 & 24.82 & 26.33 & 4.424 & 2.77\p 0.18 & 0.051\p 0.039 & 0.803 & -0.29 \\
 N227 & 12:36:44.0  & 62:12:50.1   & 19.98 & 20.61 & 21.29 & 0.557 & 2.69\p 0.12 & 0.135\p 0.004 & 0.353 & -0.09 \\
 N228 & 12:36:44.12 & 62:12:44.5 & 21.59 & 22.60 & 24.96 & 1.76 & 3.21\p 0.16 & 0.085\p 0.016 & $<$0.238 & $>$0.19 \\
 N230 & 12:36:44.4 & 62:11:33.3 & 19.33  & 20.02 & 21.75 & 1.01 & 3.77\p 0.13 & 0.107\p 0.009 & 0.774 & $<$-0.33 \\
 N231 & 12:36:44.94 & 62:11:45.3 & 27.0  & 27.0  & 27.6\tablenotemark{b} & 1.04\tablenotemark{c} & n/a & n/a & 0.623 & -0.20 \\
 N240 & 12:36:46.33 & 62:14:04.7  & 19.89 & 20.31 & 21.58 & 0.961 & 3.57\p 0.13 & 0.131\p 0.010 & 32.6 & 0.08 \\
 N245 & 12:36:47.04 & 62:12:38.2 & 20.61 & 20.82 & 21.18 & 0.321 & 3.23\p 0.14 & 0.095\p 0.013 & 0.238 & $<$0.01\\
 N249 & 12:36:48.07 & 62:13:09.0  & 19.44 & 19.99 & 20.74 & 0.475 & 3.70\p 0.18 & 0.165\p 0.011 & 2.37 & -0.40 \\
 N251 & 12:36:48.37 & 62:14:26.4 & 18.68 & 18.92 & 19.22 & 0.139 & 2.99\p 0.12 & 0.171\p 0.003 & 0.75 & $<$-0.40 \\
 N257 & 12:36:49.45 & 62:13:47.1 & 17.49 & 17.79 & 18.40 & 0.089 & 3.14\p 0.14 & 0.121\p 0.002 & 0.422 & $<$-0.10 \\
 N258 & 12:36:49.51 & 62:14:06.9  & 20.78 & 21.36 & 22.20 & 0.752 & 2.78\p 0.19 & 0.096\p 0.005 & 0.353 & $<$-0.08 \\
 N260 & 12:36:49.71 & 62:13:13.2 & 20.38 & 21.14 & 22.00 & 0.475 & 3.09\p 0.13 & 0.121\p 0.006 & 0.611 & $<$-0.39 \\
 N261 & 12:36:50.17 & 62:12:16.6 & 20.96 & 21.67 & 22.81 & 0.902 & 2.97\p 0.16 & 0.158\p 0.019 & $<$0.202 & $>$0.48 \\
 N267 & 12:36:51.73 & 62:12:21.4 & 24.01 & 25.8 & 26.03 & 2.75\tablenotemark{c} & n/a & n/a & 3.4 & 0.14 \\
 N272 & 12:36:52.76 & 62:13:54.1 & 21.51 & 21.76 & 22.16 & 1.355 & 2.71\p 0.13 & 0.637\p 0.015 & 0.334 & $<$-0.03 \\
 N274 & 12:36:52.89 & 62:14:44.1 & 18.15 & 18.64 & 19.33 & 0.321 & 3.49\p 0.12 & 0.141\p 0.005 & 2.44 & -0.62 \\
 N286 & 12:36:55.45 & 62:13:11.2 & 20.43 & 21.09 & 22.70 & 0.955 & 3.10\p 0.17 & 0.115\p 0.010 & 4.08 & -0.40\\
 N287 & 12:36:55.79 & 62:12:00.9  & 22.57 & 23.5 & 24.61 & n/a & 3.05\p 0.16 &0.517\p 0.049 & 1.26 &  -0.36  \\
 N291 & 12:36:56.62 & 62:12:45.7 & 18.91 & 19.54 & 20.49 & 0.517 & 3.58\p 0.08 & 0.086\p 0.005 & 0.378 & $<$0.03 \\
 N294 & 12:36:56.92 & 62:13:01.6  &21.05 & 21.72 & 23.28 & 0.474 & 3.44\p 0.16 & 0.128\p 0.014 & 3.14 & -0.50 \\
 N296 & 12:36:57.47 & 62:12:10.5 & 19.74 & 20.36 & 21.33 & 0.663 & 3.00\p 0.16 & 0.178\p 0.012 & 0.683 & -0.05 \\
 N353 & 12:37:8.33  & 62:10:55.9 & 19.89 & n/a & 20.28 & 0.422 & 1.064\p 0.13 & 0.141\p 0.249 & 0.984 & $<$-0.34 \\
 N460 & 12:37:42.65 & 62:07:51.7 & 20.83 & n/a & 21.76 & 1.08 & 2.65\p 0.15   & 0.214\p 0.006 & 2.31 & $<$-0.01 \\
 N463 & 12:37:45.02 & 62:07:18.9 & 24.35 & n/a & $>$27.4 & 2.50\tablenotemark{c} & n/a & n/a & 29.4 & -0.32 \\
 S226 & 3:32:36.69 & -27:46:30.8 & 21.61 & 22.21 & 23.09 & 0.766 & 2.96\p 0.16  & 0.219\p 0.005 & 1.18 & $<$-0.18\\
 S233 & 3:32:38.03 & -27:46:26.2 & 23.56 & 24.86 & 26.89 & 1.74\tablenotemark{c} & 2.66\p 0.21 & 0.123\p 0.013 & 8.76 & -0.45 \\
 S236 & 3:32:38.78 & -27:47:32.3 & 20.43 & 20.88 & 21.20 & 0.458 & 2.49\p 0.19 & 0.196\p 0.001  & 0.784 & $<$-0.13\\
 S242 & 3:32:39.08 & -27:46:01.7 & 20.44 & 20.91 & 20.86 & 1.216 & 2.88\p 0.20 & 0.162\p 0.001  & 8.06 & -0.18 \\
 S243 & 3:32:39.17 & -27:48:32.7 & 23.35 & 25.36 & 26.34 & 3.02\tablenotemark{c} & 2.51\p 0.24 & 0.219\p 0.008 & $<$0.355 & $>$0.44\\
 S245 & 3:32:39.68 & -27:48:50.7 & 22.45 & 24.06 & 24.77 & 3.064 & 2.47\p 0.19 & 0.163\p 0.004 & 7.48 & 0.18\\
 S246 & 3:32:39.73 & -27:46:11.2 & 21.86 & 23.32 & 25.25 & 1.61\tablenotemark{c} & 2.96\p 0.13 & 0.137\p 0.009 & 8.42 & -0.24 \\
 S248 & 3:32:41.5  & -27:47:17.3 & 19.73 & 20.40 & 21.17 & 0.64\tablenotemark{c} & 3.11\p 0.15 & 0.146\p 0.001 & 0.348 & $<$0.24 \\
 S254 & 3:32:42.84 & -27:47:02.4 & 24.12 & 24.45 & 24.63 & 3.193 & 2.38\p 0.23 & 0.156\p 0.016  & 6.31 & -0.33 \\
 S259 & 3:32:44.06 & -27:54:54.1 & 25.3 & $>$25.8 & 26.99 & 0.908 & n/a & n/a & 2.4 & -0.21\\
 S266 & 3:32:45.14 & -27:47:24.0 & 19.70 & 20.34 & 20.90 & 0.438 & 2.82\p 0.19 & 0.133\p 0.001 & 0.913 & $<$-0.21
\enddata
\tablenotetext{a}{10$^{-16}$ ergs s$^{-1}$ cm$^{-2}$.}
\tablenotetext{b}{Data point is for F814w filter.}
\tablenotetext{c}{Photometric redshift is used.}
\end{deluxetable}






\clearpage

\end{document}